\documentclass[usenatbib]{mn2e}
\usepackage{epsfig}
\usepackage[varg]{txfonts}
\usepackage[T1]{fontenc}

\newif\ifAMStwofonts
\AMStwofontstrue

\def\be{\begin{equation}}
\def\ee{\end{equation}}

\def\gtsima{$\; \buildrel > \over \sim \;$}
\def\ltsima{$\; \buildrel < \over \sim \;$}
\def\prosima{$\; \buildrel \propto \over \sim \;$}
\def\gsim{\lower.5ex\hbox{\gtsima}}
\def\lsim{\lower.5ex\hbox{\ltsima}}
\def\simgt{\lower.5ex\hbox{\gtsima}}
\def\simlt{\lower.5ex\hbox{\ltsima}}
\def\simpr{\lower.5ex\hbox{\prosima}}

\def\HI{\hbox{H$\,\rm \scriptstyle I\ $}}

\def\HeII{\hbox{He$\,\rm \scriptstyle II\ $}}

\title[The 21~cm forest with LOFAR and SKA]{Simulating the 21~cm forest detectable with LOFAR and SKA in the spectra of high-$z$ GRBs}

\author[B. Ciardi et al.]
{B. Ciardi$^1$\thanks{E-mail:ciardi@mpa-garching.mpg.de}, S. Inoue$^2$, F.~B. Abdalla$^{3,4}$, K. Asad$^5$, G. Bernardi$^6$, J.~S. Bolton$^7$, \and
M. Brentjens$^8$, A.~G. de Bruyn$^{5,8}$, E. Chapman$^3$, S. Daiboo$^5$, E.~R. Fernandez$^5$, \and
A. Ghosh$^5$, L. Graziani$^1$, G.~J.~A. Harker$^3$, I.~T. Iliev$^9$, V. Jeli\'{c}$^{5,8,10}$, H. Jensen$^{11}$, \and
S. Kazemi$^{12}$, L.~V.~E. Koopmans$^5$, O. Martinez$^5$, A. Maselli$^{13}$, G. Mellema$^{11}$,  \and
A.~R. Offringa$^{5,8}$, V.~N. Pandey$^5$, J. Schaye$^{14}$, R. Thomas$^5$, H. Vedantham$^5$, \and
S. Yatawatta$^8$, S. Zaroubi$^5$
 \\
$^1$ Max-Planck-Institut f\"ur Astrophysik,
     Karl-Schwarzschild-Strasse 1, D-85748 Garching b. M\"unchen, Germany\\
$^2$ Institute for Cosmic Ray Research, University of Tokyo, Tokyo, Japan\\
$^3$ University College London, Gower Street, London WC1E 6BT, UK\\
$^4$ Department of Physics \& Electronics, Rhodes University, Grahamstown 6140, South Africa\\
$^5$ Kapteyn Astronomical Institute, University of Groningen, PO Box 800, 9700 AV Groningen, the Netherlands\\
$^6$ SKA SA, 3rd Floor, The Park, Park Road, Pinelands 7405, South Africa\\
$^7$ School of Physics and Astronomy, University of Nottingham, University Park, Nottingham, NG7 2RD, UK\\
$^8$ ASTRON, PO Box 2, 7990 AA Dwingeloo, the Netherlands\\
$^9$ Astronomy Centre, Department of Physics \& Astronomy, Peven sey II Building, University of Sussex, Falmer, Brighton BN1 9QH, UK\\
$^{10}$ Ru{\dj}er Bo\v{s}kovi\'{c} Institute, Bijeni\v{c}ka cesta 54, 10000 Zagreb, Croatia\\
$^{11}$ Department of Astronomy and Oskar Klein Centre for Cosmoparticle Physics, AlbaNova, Stockholm University, SE-106 91 Stockholm, Sweden\\
$^{12}$ ASTRON \& IBM Center for Exascale technology, Oude Hoogeveensedijk 4, 7991 PD Dwingeloo, the Netherlands\\
$^{13}$ EVENT Lab for Neuroscience and Technology, Universitat de Barcelona,
     Passeig de la Vall d'Hebron 171, 08035 Barcelona, Spain\\
$^{14}$ Leiden Observatory, Leiden University, PO Box 9513, 2300RA Leiden, the Netherlands\\
}

\date{April 2015}           

\pagerange{\pageref{firstpage}--\pageref{lastpage}}
\pubyear{2014}
\begin{document}

\maketitle
\label{firstpage}

\begin{abstract}
We investigate the feasibility of detecting 21~cm absorption features in the afterglow spectra of high redshift long Gamma Ray Bursts (GRBs). 
This is done employing simulations of cosmic reionization, together with estimates of the GRB radio afterglow flux and the instrumental characteristics of the LOw Frequency ARray (LOFAR).
We find that absorption features could be marginally (with a S/N larger than a few) detected by LOFAR at $z\gsim 7$ if the GRB is a highly energetic event originating from Pop~III stars, while the detection
would be easier if the noise were reduced by one order of magnitude, i.e. similar to what is expected for the first phase of the Square Kilometer Array (SKA1-low). On the other hand, more standard GRBs are too dim to be detected even with ten times the sensitivity of SKA1-low, and only in the most optimistic case can a S/N larger than a few be reached at $z \gsim 9$.
\end{abstract}



\section{Introduction}         
\label{sec:intro}

Present and planned radio facilities such as LOFAR\footnote{http://lofar.org} \citep{vanHaarlem_etal_2013}, 
MWA\footnote{http://www.mwatelescope.org}, PAPER\footnote{http://eor.berkeley.edu} and
SKA\footnote{http://www.skatelescope.org}, aim at detecting the 21~cm signal from the Epoch of Reionization (EoR) in terms of observations such as
tomography (e.g. \citealt{Tozzi.Madau.Meiksin.Rees_2000,Ciardi.Madau_2003,Furlanetto.Sokasian.Hernquist_2004,Mellema.Iliev.Pen.Shapiro_2006,Santos_etal_2008,Geil.Wyithe_2009,Zaroubi_etal_2012,Malloy.Lidz_2013}), 
fluctuations and power spectrum (e.g. \citealt{Madau.Meiksin.Rees_1997,Shaver.Windhorst.Madau.deBruyn_1999,Tozzi.Madau.Meiksin.Rees_2000,Ciardi.Madau_2003,Furlanetto.Sokasian.Hernquist_2004,Mellema.Iliev.Pen.Shapiro_2006,Pritchard.Loeb_2008,Baek_etal_2009,Patil_etal_2014}), 
or absorption features in the spectra of high-$z$ radio-loud sources (e.g. \citealt{Carilli.Gnedin.Owen_2002,Furlanetto_2006,Xu_etal_2009,Mack.Wyithe_2012,Meiksin_2011,Xu.Ferrara.Chen_2011,Vasiliev.Shchekinov_2012,Ciardi_etal_2013,Ewall-Wice.Dillon.Mesinger.Hewitt_2014}). 
These observations would offer unique information on the statistical properties of the EoR (such as e.g. its duration), the amount of
\HI present in the intergalactic medium (IGM), and, ultimately, the history of cosmic reionization and the properties of its sources.

In particular, the detection of absorption features in 21~cm could be used to gain information on the cold, neutral hydrogen present along the line of sight (LOS) towards e.g. high-$z$ quasars, similarly to what is presently done at lower redshift with the Ly$\alpha$ forest (for a review see
\citealt{Meiksin_2009}).
While the detection and analysis of the 21~cm forest is in principle an easier task compared to imaging or even statistical measurements and could probe larger $k$-modes, these absorption like experiments are
rendered less likely to happen due to the apparent lack of high-$z$ radio-loud sources (e.g. \citealt{Carilli.Gnedin.Owen_2002,Xu_etal_2009}), the most distant being TN0924-2201 at $z$=5.19 \citep{vanBreugel_etal_1999}. 

Gamma Ray Bursts (GRBs) have been suggested as alternative background sources (e.g. \citealt{Ioka.Meszaros_2005,Inoue.Omukai.Ciardi_2007,Toma.Sakamoto.Meszaros_2011}), as they have been observed up to $z \sim 8-9$ \citep{Salvaterra_etal_2009,Tanvir_etal_2009,Cucchiara_etal_2011}, they are expected to occur up to the epoch of the first stars \citep{Bromm.Loeb_2002,Natarajan_etal_2005,Komissarov.Barkov_2010,Meszaros.Rees_2010,Campisi.Maio.Salvaterra.Ciardi_2011,Suwa.Ioka_2011,Toma.Sakamoto.Meszaros_2011} and to be visible in the IR and radio up to very high-$z$ due to cosmic time dilation effects \citep{Ciardi.Loeb_2000,Lamb.Reichart_2000}. Because of the latter, if a GRB afterglow were observed in the IR by e.g. ALMA\footnote{http://www.almaobservatory.org/} at a post-burst observer time of a few hours, this would offer a few days to plan for an observation in the radio accordingly (e.g. \citealt{Inoue.Omukai.Ciardi_2007}).

In this letter we investigate the feasibility of detecting the 21~cm forest with LOFAR and SKA in the radio afterglow of high-$z$ GRBs. In Section~2 we present the method used to calculate the forest; in Section~3 we discuss the properties of the GRBs; in Section~4 we present our results; and in Section~5 we give our conclusions.


\section{The 21~cm forest}

In this work we make use of the pipeline developed in Ciardi et al. (2013, hereafter C2013) to assess the feasibility of detecting the 21~cm forest in the spectra of high-$z$ GRBs. For this reason, here we only give a brief overview of this pipeline, while we refer the reader to the original paper for more details.

If a radio-loud source is located at redshift $z_s$, the photons emitted at a frequency $\nu>\nu_{\rm 21cm}=1.42$~GHz can be absorbed by the neutral hydrogen encountered along their LOS at $z=(\nu_{\rm 21cm}/\nu)(1 + z_s)-1$, with a probability $(1-e^{-\tau_{\rm 21cm}})$. The optical depth in the low-frequency limit, as appropriate for the 21~cm line, can be written as \cite[e.g.][]{Madau.Meiksin.Rees_1997,Furlanetto.Oh.Briggs_2006}:

\begin{eqnarray}
\label{eq:tau}
\tau_{\rm 21cm} & = & \frac{3}{32 \pi} \frac{h_p c^3 A_{\rm 21cm}}{k_B \nu_{\rm 21cm}^2}
\frac{x_{\rm HI} n_{\rm H}}{T_s (1+z) (dv_\parallel/dr_\parallel)}, 
\end{eqnarray}
where $n_{\rm H}$ is the hydrogen number density, 
$x_{\rm HI}$ is the mean neutral hydrogen fraction, 
$T_s$ is the gas spin temperature, $A_{\rm 21cm}=2.85 \times 10^{-15}$~s$^{-1}$
is the Einstein coefficient of the transition, and
$dv_\parallel/dr_\parallel$ is the gradient of the velocity along the LOS, with $r_\parallel$ comoving distance and $v_\parallel$ proper velocity including the Hubble flow and the gas peculiar velocity. The other symbols have the usual meaning.

The optical depth has been calculated using the simulation of reionization called ${\mathcal L}$4.39 in C2013. This has been obtained by post-processing a GADGET-3 \citep[an updated version of the publicly available code GADGET-2; see][]{Springel_2005} hydrodynamic simulation with the 3D Monte Carlo radiative transfer code CRASH (\citealt{Ciardi_etal_2001,Maselli.Ferrara.Ciardi_2003,Maselli.Ciardi.Kanekar_2009, Pierleoni.Maselli.Ciardi_2009,Partl_etal_2011,Graziani.Maselli.Ciardi_2013}), which follows the propagation of UV photons and evaluates self-consistently the evolution of H$\,\rm \scriptstyle I$, He$\,\rm \scriptstyle I$, \HeII and gas temperature. The hydrodynamic simulations were run in a box of size 4.39$h^{-1}$~Mpc comoving, with $2 \times 256^3$ gas and dark matter particles, and cosmological parameters $\Omega_{\Lambda}=0.74$, $\Omega_m=0.26$, $\Omega_b=0.024 h^2$, $h=0.72$, $n_s=0.95$ and $\sigma_8=0.85$, where the symbols have the usual meaning. The gas density, temperature, peculiar velocity and halo masses, were gridded onto a uniform $128^3$ grid to be processed with CRASH. The sources are assumed to have a power-law spectrum with index 3. For more details on the choice of the parameters and the results of the reionization histories we refer the reader to \cite{Ciardi.Bolton.Maselli.Graziani_2012} and C2013.
Here we further note that the simulations were designed to match WMAP observations of the Thomson scattering optical depth \citep{Komatsu_etal_2011}, while more recent Planck measurements \citep{Ade_etal_2015} favor a lower value, i.e. a delayed reionization process. This does not change our conclusions, and it actually makes them conservative, as more \HI would be expected at each redshift compared to the model considered here. 

Random LOS are cast through the simulation boxes and the corresponding 21~cm absorption is calculated.
It should be noted that here we consider a case in which the temperature of the IGM is determined only by the effect of UV photons, i.e. gas which is not reached by ionizing photons remains cold, while the effect of Ly$\alpha$ and x-ray photons on the 21~cm forest is extensively discussed in C2013.

Once the theoretical spectra are evaluated, instrumental effects and noise need to be included to calculate mock spectra. To do this we assume a radio source at redshift $z_s$, with a power-law spectrum with spectral index $\alpha$, and a flux density $S_{\rm in}(z_s)$. We then simulate spectra as they would be seen by LOFAR. 
While we refer the reader to C2013 for more details, here we just mention that the noise $\sigma_n$ is given by\footnote{We note that eq.~\ref{eq:noise} is correct as long as the $SEFD$ is calculated theoretically for a single polarization using the system temperature (as done in this paper), while when the $SEFD$ is determined observationally from Stokes I the noise is reduced by a factor of sqrt(2), since it combines the two cross-dipole sensitivities.}:

\begin{equation}
{\sigma _n} = \frac{W}{{{n_s}}}\frac{{SEFD}}{{\sqrt {2N\left( {N - 1} \right)
{t_{{\mathop{\rm int}} }} \Delta \nu} }},
\label{eq:noise}
\end{equation}
where $W\sim 1.3$ is a factor which incorporates the effect of weighting, $n_s=0.5$ is the system efficiency, $\Delta\nu$ is the bandwidth, $t_{int}$ is
the integration time, $N=48$ is the number of stations, and the system equivalent flux density is given by:

\begin{equation}
SEFD = \frac{{2{\kappa _B}{T_{\rm sys}}}}{{N_{\rm dip}{\eta _\alpha }{A_{\rm eff}}}},
\label{eq:sefd}
\end{equation}
where $\kappa _B$ is Boltzmann's constant, $A_{\rm
eff}=min(\frac{\lambda^2}{3}, 1.5626)$~m$^2$ is the effective area of each dipole\footnote{Note that the same expression in C2013 contains a typo.}, $N_{\rm
dip}$ is the number of dipoles per station (we assume 24 tiles per station with 16 dipoles each),
$\eta_{\alpha}=1$ is the dipole efficiency, and the system noise is $T_{\rm sys}=[140+60(\nu/300 \; {\rm MHz})^{-2.55}]$~K\footnote{
We note that the values obtained from eq.~\ref{eq:sefd} are very similar to the real ones reported in the LOFAR official webpage.}.
As a reference, $\sigma_n=0.66$~mJy for $\nu=130$~MHz, $t_{\rm int}=1000$~h and $\Delta \nu=10$~kHz.


\section{Radio afterglows of high-z GRBs}
\label{sec:grbs}

Here we discuss aspects of the radio afterglow emission from GRBs
\cite[see e.g.][for recent reviews]{vanEerten_2013,Granot.Horst_2014}
that are most relevant for studies of the 21 cm forest at high redshift.
GRB afterglows consist primarily of broadband synchrotron emission from nonthermal electrons accelerated
in the forward shock of relativistic blastwaves, which are driven into the ambient medium by
transient, collimated jets triggered by the GRB central engine.
The low-frequency radio flux is typically suppressed in the beginning due to synchrotron self-absorption
and rises gradually as the blastwave expands.
The light curve at a given frequency $\nu$ peaks when the emission becomes optically thin to self-absorption,
after which it decays, according to the overall behavior of the decelerating blastwave.
As background sources for observing the 21 cm forest, the emission near this peak flux time is naturally the most interesting.

The expected observer peak time, $t_{pk}(\nu)$, and the corresponding peak flux, $S_{\rm in}(\nu)$, can be evaluated
using the formulation outlined in the Appendix of \citet{Inoue_2004}, which is based on standard discussions in the literature
\cite[e.g.][]{Sari.Piran.Narayan_1998,Sari.Piran.Halpern_1999,Panaitescu.Kumar_2000} and is sufficient for our purposes of
estimating the radio afterglow emission at relatively late times after the burst.
We consider cases for which the peak time occurs after the crossing time of the minimum frequency, $\nu_m$, as well as the jet break time but before the non-relativistic transition time, valid for the range of parameters chosen here\footnote{Although the time evolution after the jet break is not explicitly described in \citet{Inoue_2004}, it is taken into account following \citet{Sari.Piran.Halpern_1999}.}.
For concreteness, the spectral index of the accelerated electron distribution is taken to be $p=2.2$,
implying a radio spectral index of $\alpha=0.6$ for the optically thin, slow cooling regime.
We also choose $\epsilon_e=0.1$ and $\epsilon_B=0.01$ for the fractions of post-shock energy
imparted to accelerated electrons and magnetic fields, respectively,
consistent with observationally inferred values \cite[][]{Panaitescu.Kumar_2001}.
This gives:
\begin{eqnarray}
t_{pk}(\nu) &\simeq& 540 \ {\rm days} \ E_{53}^{0.44} n_0^{0.2} \ \theta_{-1}^{0.88} \\ \nonumber
                  &\times& \left(1 + z \over 11 \right)^{0.03} \left(\nu \over 10^8 \  {\rm Hz} \right)^{-0.97}
\label{eq:peaktime}
\end{eqnarray}
and
\begin{eqnarray}
S_{\rm in}(\nu) &\simeq& 1.1 \times 10^{-3} \ {\rm mJy} \ E_{53}^{0.77} n_0^{-0.38} \ \theta_{-1}^{1.54} \\ \nonumber
                  &\times& \left(D_{\rm L}(z) \over 3.3 \times 10^{29} \ {\rm cm}\right)^{-2} \left(1 + z \over 11 \right)^{2.53} \left(\nu \over 10^8 \ {\rm Hz}\right)^{-1.53}
\label{eq:peakflux}
\end{eqnarray}
where $E=10^{53} E_{53}$ erg is the isotropic-equivalent blastwave kinetic energy,
$\theta=0.1 \ \theta_{-1}$ rad is the jet collimation half-angle,
$n_{medium}=n_0 \ {\rm cm^{-3}}$ is the ambient medium number density, and $D_L(z)$ is the luminosity distance.

The known population of GRBs (referred to as GRBII, as they are expected to originate from standard Pop~II/I stars) has been observationally inferred to possess
values of these quantities up to $E \sim 10^{54}$ erg, $\theta \sim 0.3$ rad and/or
down to $n_{medium} \sim 10^{-4} {\rm cm^{-3}}$ \cite[][]{Panaitescu.Kumar_2001,Ghirlanda_etal_2013a}.
Thus, they may provide fluxes up to $S_{\rm in} \sim$ 0.1 mJy at $\nu \sim$100 MHz and $t_{pk} \sim$ 3000 days
(i.e. as sources virtually steady over several years) for events at $z \sim 10$ under favorable conditions \cite[see also][]{Ioka.Meszaros_2005,Inoue.Omukai.Ciardi_2007}.

On the other hand, although not yet confirmed by observations, an intriguing possibility for high-$z$ detections is the existence of GRBs arising
from Population III stars (GRBIII), with significantly larger blastwave energies, up to values as high as $E \sim 10^{57}$ erg,
by virtue of their prolonged energy release fueled by accretion of the extensive envelopes of their progenitor stars
\cite[][]{Meszaros.Rees_2010,Komissarov.Barkov_2000,Suwa.Ioka_2011}.
Compared to known GRBs, their blastwaves can expand to much larger radii and consequently with much brighter low-frequency
radio emission, potentially exceeding $S_{\rm in} \sim$10 mJy at $\nu \sim$100 MHz and $t_{pk} \sim 3 \times 10^4$ days
for events at $z \sim 20$ \citep{Toma.Sakamoto.Meszaros_2011,Ghirlanda_etal_2013b}.

Note that although more recent and detailed theoretical studies of afterglow emission relying on hydrodynamical simulations
have revealed non-trivial deviations from the simple description presented above \cite[][]{vanEerten_2013,Ghirlanda_etal_2013b},
the latter should still be sufficient for our aim of discussing expectations for observations of 21~cm forest.                                    

Finally, as reference numbers, \cite{Campisi.Maio.Salvaterra.Ciardi_2011} find that $\sim$ 1 ($<0.06$) yr$^{-1}$~sr$^{-1}$ GRBII (GRBIII) are predicted to lie at $z>6$. This translates into $\sim 7.5 \times 10^{-3}$ GRBII ($\sim 4.5 \times 10^{-4}$ GRBIII) per year in a 25~deg$^2$ (LOFAR) field of view, and $\sim 3$ ($\sim 0.2$) per year in a 10$^4$~deg$^2$ (SKA) field of view.


\begin{figure}
\centering
\includegraphics[width=0.5\textwidth]{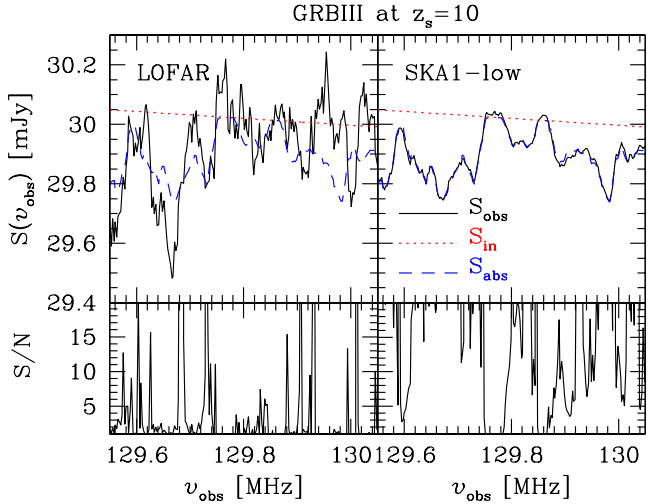}
\caption{{\it Upper panels:} Spectrum of a GRBIII positioned at
   $z_s=10$ (i.e. $\nu \sim 129$~MHz), with a flux density $S_{\rm in}(z_s)=30$~mJy.  The red
   dotted lines refer to the intrinsic spectrum of the source,
   $S_{\rm in}$; the blue dashed lines to the simulated spectrum for
   21~cm absorption, $S_{\rm abs}$; and the black solid lines to the
   spectrum for 21~cm absorption as it would be seen with a bandwidth $\Delta \nu=10$~kHz, 
   after an integration time $t_{int}=1000$~h. The left and right panels
   refer to a case with the noise $\sigma_n$ given in eq.~\ref{eq:noise} (LOFAR telescope) and
   with 0.1$n$ (expected for SKA1-low), respectively.
   {\it Lower panels:} S/N corresponding to the upper panels.
   See text for further details.  
}
\label{fig:GRBIII_z10}
\end{figure}

\section{Results}
\label{sec:results}

For an easier comparison to a case in which the background source is a radio-loud QSO, here we show results for the same LOS and at the same redshift of C2013 (see their Figs.~11, 12 and 13).

In the upper panels of Figure~\ref{fig:GRBIII_z10} we plot the spectrum of a GRBIII at $z_s=10$ (i.e. $\nu \sim 129$~MHz) with a flux density $S_{\rm in}(z_s)=30$~mJy. The simulated absorption spectrum, $S_{\rm abs}$, is shown together with the observed spectrum, $S_{\rm obs}$, calculated assuming an observation time $t_{int}=1000$~h, a bandwidth $\Delta \nu=10$~kHz and a noise $\sigma_n$ (given in eq.~\ref{eq:noise}; left panels) and 0.1~$\sigma_n$ (similar to the value expected for SKA1-low\footnote{SKA1-low is the first phase of the SKA covering the lowest frequency band.}, which is $\sim$1/8th of the LOFAR noise; right panels).  
In the lower panels of the Figure we also show the quantity $\left| S_{\rm in}-S_{\rm abs}\right|/ \left|S_{\rm obs}-S_{\rm abs}\right|$, which effectively represents the signal-to-noise with which we would be able to detect the absorption. If indeed such powerful GRBs exist, then absorption features could be detected by LOFAR with an average\footnote{We note that the definition of `average' is somewhat arbitrary and depends on the frequency range used. Here the average refers to the one calculated over the frequency range shown in the Figures.}  signal-to-noise S/N$\sim$5, while if the noise were reduced by a factor of 10 the detection would be much easier.  

\begin{figure}
\centering
\includegraphics[width=0.5\textwidth]{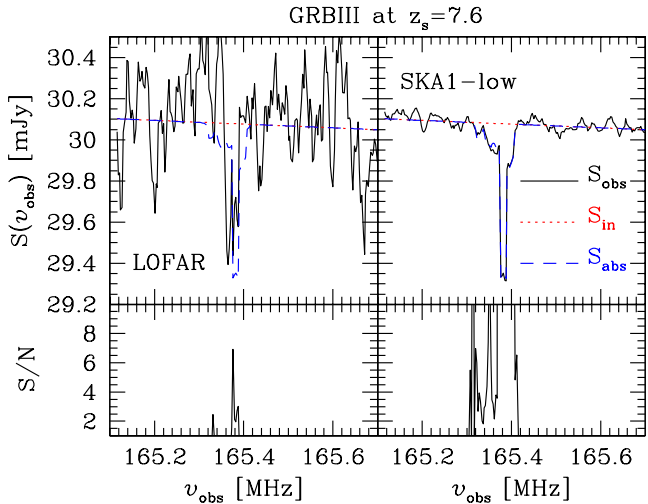}
\caption{As Figure~\ref{fig:GRBIII_z10}, but the GRBIII is positioned at $z_s=7.6$ (i.e. $\nu \sim 165$~MHz) and $\Delta \nu=5$~kHz.}
\label{fig:GRBIII_z7.6}                 
\end{figure}

Figure~\ref{fig:GRBIII_z7.6} shows a LOS to the same GRB located at $z=7.6$, when the IGM is $\sim 80\%$ ionized by volume. The LOS has been chosen to intercept a pocket of gas with $\tau_{\rm 21cm}>0.1$ to show that strong absorption features could be detected, albeit by LOFAR only marginally with a S/N of a few, also at a redshift when most of the IGM is in a highly ionized state.

\begin{figure}
\centering
\includegraphics[width=0.5\textwidth]{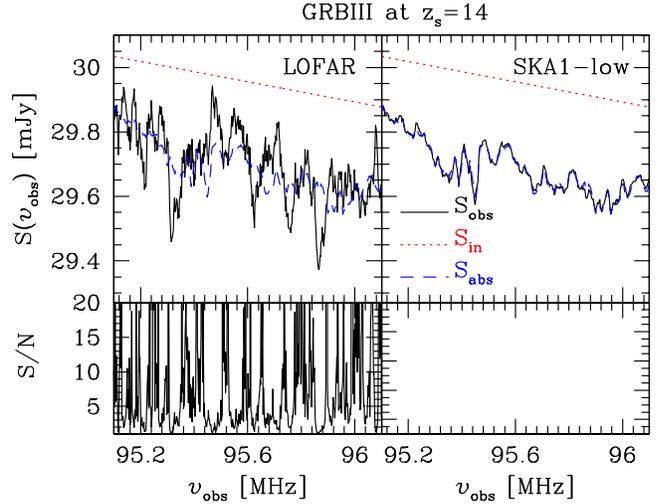}
\caption{As Figure~\ref{fig:GRBIII_z10}, but the GRBIII is positioned at $z_s=14$ (i.e. $\nu \sim 95$~MHz) and $\Delta \nu=20$~kHz. Note that the S/N in the lower right panel is always higher than the range covered by the axis.}
\label{fig:GRBIII_z14}
\end{figure}

On the other hand, a strong average absorption (rather than the absorption features seen in the previous figures) can be easily detected (S/N $>10$) in the spectra of GRBs located at very high redshift, as shown in Figure~\ref{fig:GRBIII_z14}. If the intrinsic spectrum of the source could be inferred through other means, for example, accurate spectral measurements of the unabsorbed continuum at GHz frequencies and above, such detection could be used to infer the global amount of neutral hydrogen present in the IGM.

\begin{figure}
\centering
\includegraphics[width=0.5\textwidth]{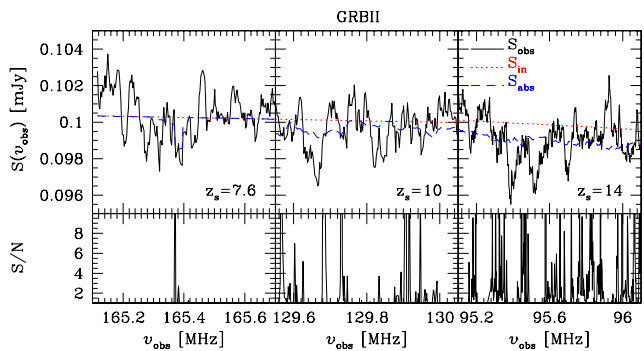}
\caption{{\it Upper panels:} Spectrum of a GRBII with a flux density $S_{\rm in}(z_s)=0.1$~mJy. The red dotted lines refer to the intrinsic spectrum of the source, $S_{\rm in}$; the blue dashed lines to the simulated spectrum for 21~cm absorption, $S_{\rm abs}$; and the black solid lines to the spectrum for 21~cm absorption as it would be seen after an integration time $t_{int}=1000$~h with a noise 0.01~$\sigma_n$ (i.e. 1/10th of the SKA1-low noise). The panels refer to a case with $z_s=7.6$ and $\Delta \nu=5$~kHz (left), $z_s=10$ and $\Delta \nu=10$~kHz (middle), $z_s=14$ and $\Delta \nu=20$~kHz (right).
   {\it Lower panels:} S/N corresponding to the upper panels.
   See text for further details.  }
\label{fig:GRBII}
\end{figure}

We have applied the same pipeline also to more standard GRBII, which have a flux density two to three orders of magnitude lower than a GRBIII. In this case we find that, even if we could collect 1000~h of observations with a noise 0.01~$\sigma_n$ (i.e. 1/10th of the SKA1-low noise), these would be barely enough to detect absorption features in 21~cm.
Also in the most optimistic case, with $S_{\rm in}(z_s)=0.1$~mJy, a positive detection would be extremely difficult at any redshift, as shown in Figure~\ref{fig:GRBII}, and only at $z \gsim 9$ a S/N larger than a few could be reached.

\section{Conclusions}

In this letter we have discussed the feasibility of detecting 21~cm absorption features in the spectra of high redshift GRBs with LOFAR and SKA. The distribution of H$\,\rm \scriptstyle I$, gas temperature and velocity field used to calculate the optical depth to the 21~cm line have been obtained from the simulations of reionization presented in \cite{Ciardi.Bolton.Maselli.Graziani_2012} and C2013. 
We find that:
\begin{itemize}
\item absorption features in the spectra of highly energetic GRBs from Pop~III stars could be marginally (with a S/N larger than a few, depending on redshift) detected by LOFAR at $z \gsim 7$;
\item the same features could be easily detected if the noise were reduced by one order of magnitude (similar to what is expected for SKA1-low);
\item the flux density of a more standard GRB is too low for absorption features to be detected even with ten times the sensitivity of SKA1-low. Only in the most optimistic case with a flux density of 0.1~mJy, can a S/N larger than a few be reached at $z \gsim 9$.
\end{itemize}
The problem of a small flux could be alleviated in case of a lensed GRB. Lensing of high-$z$ sources has been already discussed in the literature both from a theoretical (e.g. \citealt{Wyithe.Yan.Windhorst.Mao_2011}), and an observational (e.g. with the Frontier Fields as in \citealt{Oesch_etal_2015}) perspective.
Alternatively, a statistical detection of the 21~cm forest could be attempted, as already suggested by several authors (e.g. \citealt{Meiksin_2011,Mack.Wyithe_2012,Ewall-Wice.Dillon.Mesinger.Hewitt_2014}).

\section*{Acknowledgments}
The authors thank an anonimous referee for his/her useful comments.
BC acknowledges Benoit Semelin for interesting discussions.
This work was supported by DFG Priority Programs 1177 and 1573.  
SI appreciates support from Grant-in-Aid for Scientific Research No. 24340048 from MEXT of Japan.
GH has received funding from the People Programme (Marie Curie Actions) of the European Union's Seventh Framework Programme (FP7/2007--2013) under REA grant agreement no.\ 327999.
LVEK, HV, KA and AG acknowledge the financial support from the European
Research Council under ERC-Starting Grant FIRSTLIGHT - 258942.
ITI was supported by the Science and Technology Facilities Council [grant number ST/L000652/1].
VJ would like to thank the Netherlands Foundation for Scientific Research (NWO) for financial support through VENI grant 639.041.336.
JSB acknowledges the support of a Royal Society University Research Fellowship.

\bibliographystyle{apj} 
\bibliography{21cmForest_GRB}

\label{lastpage}

\end{document}